\documentstyle[aps,prl,multicol,epsf]{revtex}

\begin{document}

\title{Intrinsic tunneling spectroscopy:
A look from the inside at HTSC.}

\author{V.M.Krasnov}

\address{Department of Microtechnology and Nanoscience, Chalmers
University of Technology, S-41296 G\"oteborg, Sweden}

\date{\today }
\maketitle

\begin{abstract}

Layered structure of Bi-2212 high $T_c$ superconductor (HTSC),
provides a unique opportunity to probe quasiparticle density of
states inside a bulk single crystal by means of intrinsic
(interlayer) tunneling spectroscopy. Here I present a systematic
study of intrinsic tunneling characteristics of Bi-2212 as a
function of doping, temperature, magnetic field and intercalation.
An improved resolution made it possible to simultaneously trace
the superconducting gap (SG) and the normal state pseudo-gap (PG)
in a close vicinity of $T_c$ and to analyze closing of the PG at
$T^*$. The obtained doping phase diagram exhibits a critical
doping point for appearance of the PG and a characteristic
crossing of the SG and the PG close to the optimal doping. All
this points towards coexistence of two different and competing
order parameters in Bi-2212.



\end{abstract}

\begin{multicols}{2}

\section{Introduction}

Existence of a pseudo-gap (PG) in the electronic density of states
(DOS) of High $T_c$ Superconductors (HTSC) at temperatures above
the superconducting critical temperature, $T>T_c$ remains one of
the main challenges for understanding the HTSC phenomenon.
Theoretical explanations of the PG can be divided into
superconducting or non-superconducting classes. A pre-formed pair
scenario emphasizes that a (very) strong electron coupling
interaction may cause formation of pre-formed electron pairs at
$T$ well above $T_c$, which undergo Bose-Einstein condensation at
$T_c$, similar to a superfluid transition in a liquid He
\cite{preformed}. A precursor superconductivity scenario
emphasizes the strength of phase fluctuations in HTSC due to a
small coherence length, low density of charge carriers, quasi-2D
structure and high temperature. Strong phase fluctuations can
cause a destruction of macroscopic phase coherence at $T$ well
below the mean-field $T_{c0}$, by a mechanism similar to the
Kosterlitz-Thouless transition \cite{fluctuation}. The PG
temperature, $T^*$ is then associated with the mean-field
$T_{c0}$, which is higher than a true $T_c$ at which the
macroscopic phase coherence is established. The amplitude
fluctuations of the superconducting order parameter at $T_c$ are
supposed to be relatively small, in contrast to conventional
low-$T_c$ superconductors. Recently it was claimed that
vortex-like excitations  exist at $T$ several times larger than
$T_c$, supporting the phase-fluctuation scenario
\cite{vortexfluct}. However, phase-fluctuation scenario alone puts
more question marks than answers as it would need to explain the
$T_{c0}$ in the range of 1000 K in underdoped HTSC.

Alternatively, several non-superconducting scenarios of the PG
were proposed. According to those the PG is associated with an
additional order parameter, such as charge, spin\cite{CDSDW} or
d-density\cite{dDW} waves, independent and competing with the
superconducting order parameter. This idea is substantiated by a
reasonable nesting of the Fermi surface in HTSC. Finally,
charge-spin ordering may cause formation of one-dimensional
metallic stripes, which may be favorable for appearance of HTSC
\cite{Stripe}. In this case three characteristic temperatures may
exist: the temperature of stripe formation, $T_{c0}$ at the stripe
and the $T_c$ at which the phase coherence between stripes is
achieved.

Unfortunately, discrimination between those distinct scenarios is
restricted by the lack of consensus in experimental data, obtained
by different techniques. Various direct spectroscopic techniques
provide conflicting results. For example, ARPES indicated that the
energy gap (PG) vanishes at $\sim$ 100-110 K in optimally doped
Bi-2212, i.e., $\sim$ 15 K above $T_c$ \cite{ARPES}; break
junction technique reveals a significant temperature dependence of
the gap at $T<T_c$ but does not show any PG at $T>T_c$ in
optimally doped Bi-2212 \cite{Zasad}; while the surface tunneling
shows literally no temperature dependence of the energy gap and
persistence of the PG up to almost room temperature \cite{Renner}.

The present state of confusion requires further studies using
advanced experimental techniques. One of those is an interlayer
tunneling spectroscopy, which is unique in it's ability to measure
properties {\it inside} HTSC single crystals. This method is
specific to strongly anisotropic HTSC, like
Bi$_2$Sr$_2$CaCu$_2$O$_{8+\delta}$ (Bi-2212), in which mobile
charge carriers are localized in double CuO$_2$ layers, while the
transverse ($c-$axis) transport is due to interlayer tunneling
\cite{Kleiner,Fiske}. Interlayer tunneling has become a powerful
tool for studying both electron\cite{KrTemp,KrMag,Suzuki} and
phonon\cite{Schlenga} DOS of HTSC. It has several important
advantages compared to surface tunneling techniques: (i) it probes
bulk properties and is insensitive to surface deterioration or
surface states; (ii) the current direction is well defined; (iii)
the tunnel barrier is atomically perfect and has no extrinsic
scattering centers; (iv) mesa structures are mechanically stable
and can sustain high bias

\begin{figure}
\noindent
\begin{minipage}{0.48\textwidth}
\epsfxsize=0.8\hsize \centerline{ \epsfbox{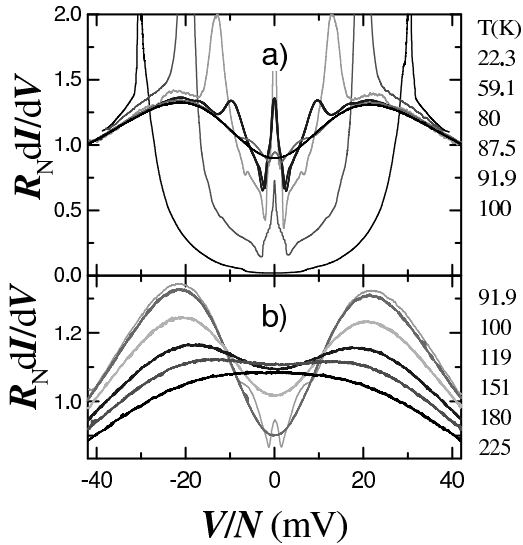} }
\vspace*{6pt} \caption{ d$I$/d$V$($V$) curves for a slightly OD
sample $T_c = $ 93 K: a) below and just above $T_c$, b) just below
and above $T_c$.} \label{Fig.1}
\end{minipage}
\end{figure}

\noindent in a wide range of temperatures ($T$) and magnetic
fields ($H$).

Here I present a systematic study of intrinsic tunneling
characteristics of Bi-2212 as a function of doping, temperature,
magnetic field and intercalation. An improved resolution made it
possible to simultaneously trace the superconducting gap (SG) and
the normal state pseudo-gap (PG) in a close vicinity of $T_c$ and
to analyze closing of the PG at $T^*$. The obtained doping phase
diagram exhibits a critical doping point for appearance of the PG
and a characteristic crossing of the SG and the PG close to the
optimal doping. This points towards coexistence of two different
and competing order parameters in Bi-2212. Finally, numerical
simulations of SIS tunneling characteristics in the perpendicular
magnetic field are performed using a circular cell approximation.
It is shown that a spatial non-uniformity due to presence of
Abrikosov vortices may cause a divergence between experimentally
measurable characteristics of the DOS. For example, the maximum in
the spatial-averaged DOS, probed in SIN tunneling experiments, can
have a qualitatively different magnetic field dependence than the
maximum superconducting gap and can be considerably different from
the peak in the differential conductance of the SIS junction.

\section{Temperature and Doping dependencies}

Small ($2-5 \mu m$ in the ab-plane) mesa structures were made on
top of Bi-2212 single crystals by a microfabrication technique.
Mesas typically contained 7-12 intrinsic Josephson junctions in
series. Measurements were performed in a three probe
configuration. A nonlinear contact resistance was approximately
equal to the resistance

\begin{figure}
\noindent
\begin{minipage}{0.48\textwidth}
\epsfxsize=0.8\hsize \centerline{ \epsfbox{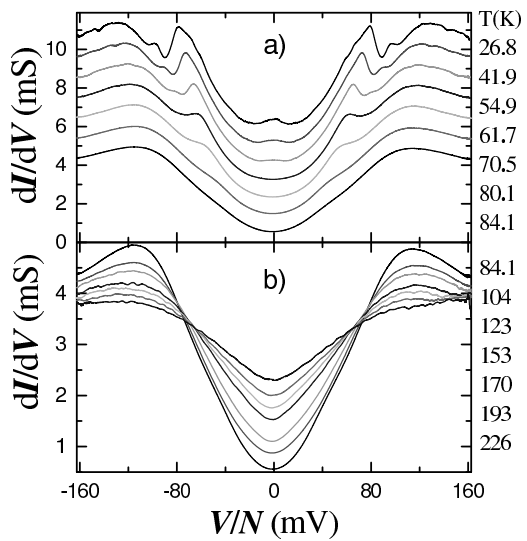} }
\vspace*{6pt} \caption{d$I$/d$V$ curves for the UD sample
$T_c=84.4 K$: a) below $T_c$ (curves are shifted for clarity), b)
above $T_c$} \label{Fig.2}
\end{minipage}
\end{figure}

\noindent of one intrinsic junction, most probably caused by a
suppression of superconductivity in the top CuO layer due to a
proximity effect with the normal electrode and surface
deterioration. For details of sample fabrication and measurements
see Ref.\cite{KrTemp}. Effect of self-heating in such small mesas
was shown to be of a minor significance \cite{Heat}, thus
contributing to an improved spectroscopic resolution.

In Figs. 1 and 2, experimental dI/dV curves for slightly overdoped
(OD) and underdoped (UD) samples are shown. Below $T_c$ a sharp
peak, corresponding to the knee in IVC's, is seen. The peak
voltage, $V_{peak}$, decreases as $T \rightarrow T_c$. Above $T_c$
the peak disappears (therefore we associate the peak with the
superconducting gap (SG)), but a distinctly different dip-and-hump
structure remains, representing the persisting PG \cite{KrTemp}.
For OD samples, $V_{peak}$ can be clearly traced up to $T_c$ and
$V_{peak} \rightarrow 0$ at $T_c$, see Figs. 1 a). At $\sim 150
K$, $V_{hump}$ starts to decrease and vanishes at $T^* \sim 200 K$
\cite{Doping}. Interestingly, IVC's are nonlinear even above
$T^*$, see Fig. 1 b). The $\sigma(V)$ curves have an inverted
parabolic shape, which might indicate the presence of van-Hove
singularity close to the Fermi level in slightly OD samples. The
behaviour of the SG in UD samples at $T \rightarrow T_c$ is one of
the most important and yet controversial issues
\cite{KrTemp,Fisher}. For UD samples the peak is much weaker than
for OD samples even at low $T$, cf. Figs. 1 a) and 2 a), and it
rapidly smears out with increasing $T$.

Remarkably for the UD sample the hump is clearly observable at any
$T$. Below $T_c$ the PG hump and the SG peak coexist and shift
simultaneously to higher voltages with decreasing $T$ (for more
details about doping and temperature dependence of the SG and PG
see Ref.\cite{Doping}). Furthermore, it is seen that $V_{peak}(T
\ll T_c)$ is

\begin{figure}
\noindent
\begin{minipage}{0.48\textwidth}
\epsfxsize=0.8\hsize \centerline{ \epsfbox{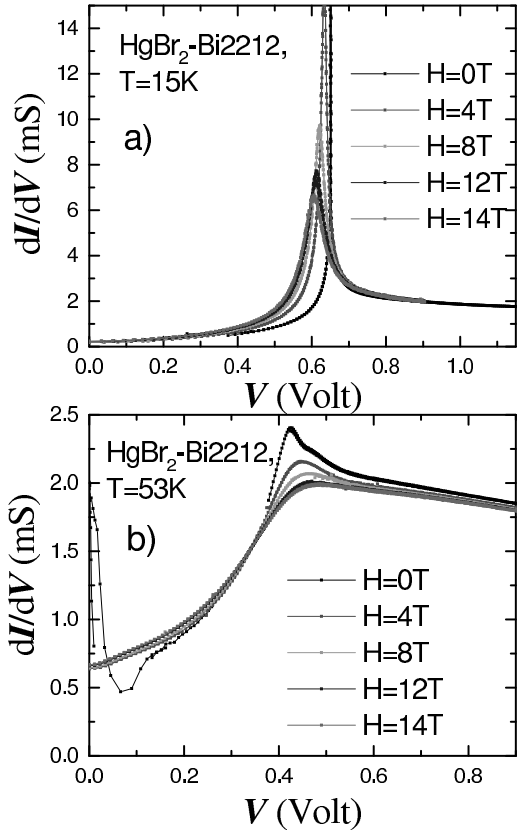} }
\vspace*{6pt} \caption{d$I$/d$V$ curves for an intercalated sample
at different magnetic fields along the c-axis at a) $T=15K$ and b)
$T=53K$} \label{Fig.3}
\end{minipage}
\end{figure}

\noindent substantially larger than the hump voltage
$V_{hump}(T_c)$ in the OD mesa (Fig.1), substantially smaller in
the UD mesa (Fig.2), while they are approximately equal in
optimally doped samples \cite{KrTemp}. Therefore, at the doping
phase diagram the Pseudo-gap line {\it crosses} the
superconducting  gap line near optimal doping and indicates the
presence of the critical doping point on the overdoped side at
$p\simeq 0.19$, as demonstrated in Ref.\cite{Doping}.

\section{Magnetic field dependency}

A response to magnetic field ($H$) provides a crucial test for the
superconducting origin of the two gaps observed in intrinsic
tunneling experiments \cite{KrMag}. Both $T$ and $H$ are depairing
parameters and suppress superconductivity when exceeding $T_c$ or
the upper critical field $H_{c2}$, respectively. However, they
have a possibility to act differently on the two gaps. Namely,
unlike $T$, $H$ may selectively affect the SG.

In Fig.3 magnetic field dependencies of dI/dV curves for an
intercalated HgBr$_2$-Bi2212 sample with $T_c \simeq 73 K$ are
shown at a) $T=15 K$ and b) $T=53 K$. Two effects should be
recognized: (i) the zero bias conductance increases (negative
magnetoresistance) roughly linearly

\begin{figure}
\noindent
\begin{minipage}{0.48\textwidth}
\epsfxsize=0.8\hsize \centerline{ \epsfbox{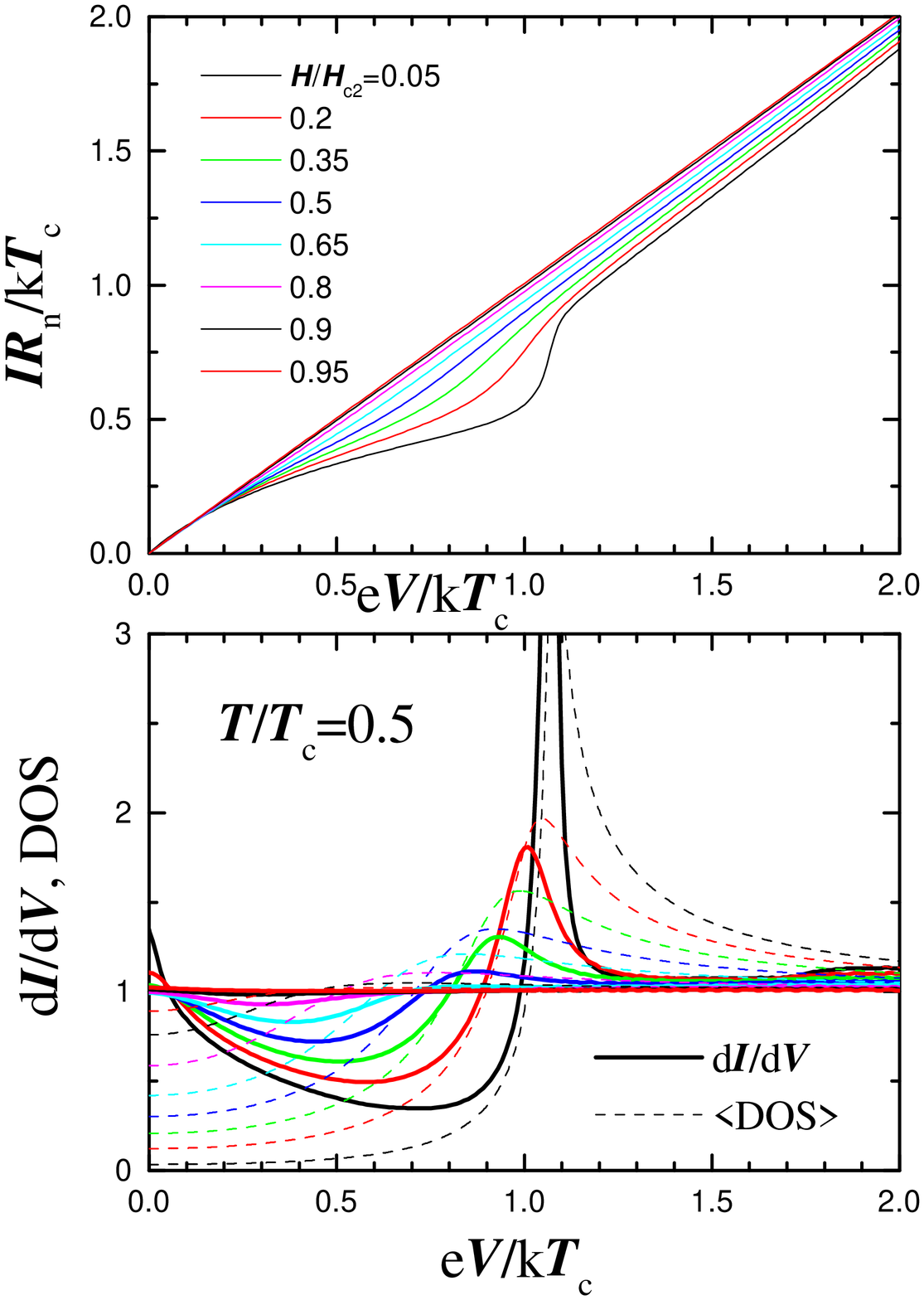} }
\vspace*{6pt} \caption{Simulated magnetic field dependencies for a
conventional s-wave SIS junction at $T=0.5 T_c$: a) IVC's, b)
dI/dV - solid lines and spatially averaged density of states
-dashed lines. } \label{Fig.4}
\end{minipage}
\end{figure}

\noindent with field (see Ref. \cite{KrMag} for more details).
This is typical for SIS junctions in which the linear increase of
conductance with $H$ is due to a linear increase of the amount of
vortices in S-electrodes, see inset in Fig. 5. (ii) It is seen
that at low $T$ the superconducting peak in dI/dV is strongly
suppressed and is shifted towards lower voltages with magnetic
field. This is also typical for SIS junctions and is caused by a
suppression of the maximum SG in the presence of the vortex
lattice. At higher $T$ the peak can be suppressed completely in 14
T (see Ref. \cite{KrMag}) and is shifted slightly outwards with
field. Such behavior is some-what less transparent, however is
also fully consistent with the behavior of SIS junctions. Indeed
numerical simulations clearly demonstrate that at elevated
temperatures the peak in spatially averaged DOS is smeared out and
moves outwards with field, see Fig. 18 in Ref.\cite{Ichioka}.

To get a deeper insight into the magnetic field dependence of the
SIS junction, numerical simulations for a conventional SIS
junction with s-wave order parameter has been performed within a
circular cell approximation \cite{GolubH}. Results of simulation
are shown in Figs. 4 and 5. In this model an explicit analytic
solution for the DOS is available at $H_{c2}-H \ll H_{c2}$. As $H
\rightarrow H_{c2}$ the superconducting order parameter $\Delta
\rightarrow 0$, as seen in Fig. 5. On the contrary, the energy of
the peak in spatially averaged DOS first saturates and then
increases with $H/H_{c2}$ \cite{GolubH}. As shown in Fig. 5, even
at very low $T$ SIN tunneling,

\begin{figure}
\noindent
\begin{minipage}{0.48\textwidth}
\epsfxsize=0.8\hsize \centerline{ \epsfbox{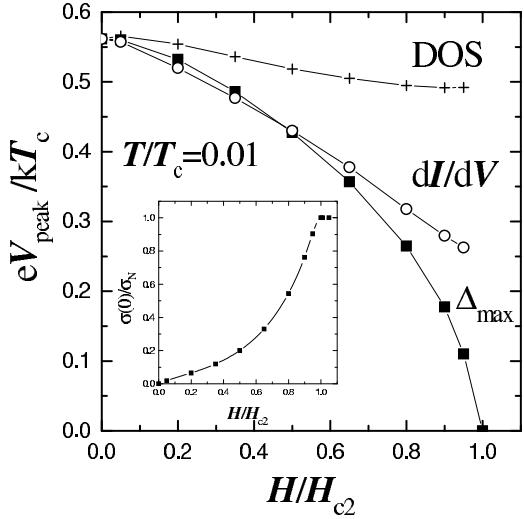} }
\vspace*{6pt}\caption{Simulated magnetic field dependencies of the
peak in spatially averaged DOS (crosses), peak in dI/dV of an SIS
junction (circles), and maximum superconducting order parameter
$\Delta_{max}$ at $T=0.01 T_c$. Inset shows magnetic field
dependence of the zero bias conductance.} \label{Fig.5}
\end{minipage}
\end{figure}

\noindent which senses the spatially averaged DOS, does not
provide clear information about the superconducting gap. The peak
in SIS tunneling characteristics follows much closer the
superconducting gap, however even SIS fails to provide clear
information at $H \rightarrow H_{c2}$. This has to be taken into
account when analyzing experimental data in strong magnetic
fields, obtained by different experimental techniques.

In conclusion, temperature\cite{KrTemp}, magnetic
field\cite{KrMag} and doping\cite{Doping} dependent intrinsic
tunneling spectroscopy provides strong evidence for independent
and competing origins of the superconductivity and the pseudo-gap
phenomenon in HTSC. This is supported by observation of (i)
coexistence of the pseudo-gap and the superconducting gap at
$T<T_c$; (ii) correlated temperature dependence of the PG hump and
the SG peak at $T<T_c$; (iii) closing the SG at $T \rightarrow
T_c$ and $H \rightarrow H_{c2}$; (iv) different magnetic field
dependence of the SG and the PG; (v) crossing of the SG and the PG
at the doping phase diagram and indication for the existence of
the critical doping point.

Acknowledgements. Financial support from the Swedish Research
Council (VR 621-2001-3236) and Knut and Alice Wallenberg
foundation are gratefully acknowledged. I am grateful to A.Golubov
for an assistance in numerical simulations.

\end{multicols}

\end{document}